\documentclass[12pt]{spieman}  
\usepackage{amsmath,amsfonts,amssymb}
\usepackage{graphicx}
\usepackage{setspace}
\usepackage{tocloft}

\title{Development of a blood oxygenation phantom for photoacoustic tomography combined with online pO$_{\textbf{2}}$ detection and flow spectrometry}

\author[a,b]{Marcel Gehrung}
\author[a,b]{Sarah E. Bohndiek}
\author[a,b,*]{Joanna Brunker}
\affil[a]{Cancer Research UK Cambridge Institute, Li Ka-Shing Centre, Cambridge, UK}
\affil[b]{Department of Physics, University of Cambridge, Cambridge, UK}

\cftpagenumbersoff{figure}
\cftpagenumbersoff{table} 

\usepackage{xcolor}
\usepackage{siunitx}
\DeclareSIUnit{\molar}{M}
\def\muarl{\ensuremath {\mu_a(\mathbf{r},\lambda)}}
\def\musrl{\ensuremath{\mu'_s(\mathbf{r},\lambda)}}
\def\Prl{P(\mathbf{r},\lambda)}
\def\Hrl{H(\mathbf{r},\lambda)}
\def\Phirl{\Phi[\mathbf{r},\lambda,\muarl,\musrl]}

\begin{document} 
\sisetup{range-phrase=--}
\maketitle

\begin{abstract}
Photoacoustic tomography (PAT) is intrinsically sensitive to blood oxygen saturation (sO$_2$) \textit{in vivo}. However, making accurate sO$_2$ measurements without knowledge of tissue- and instrumentation-related correction factors is extremely challenging. We have developed a low-cost flow phantom system to facilitate validation of photoacoustic tomography systems. The phantom is composed of a flow circuit, which is partially embedded within a tissue mimicking phantom, with independent sensors providing online monitoring of the optical absorption spectrum and partial pressure of oxygen in the tube. We first establish the flow phantom using two small molecule dyes that are frequently used for photoacoustic imaging: methylene blue (MB) and indocyanine green (ICG). We then demonstrate the potential of the phantom for evaluating sO$_2$ using chemical oxygenation and deoxygenation of blood in the phantom. Using this dynamic assessment of the photoacoustic sO$_2$ measurement in phantoms in relation to a ground truth, we explore the influence of multispectral processing and spectral coloring on accurate assessment of sO$_2$. Future studies could exploit this low-cost dynamic flow phantom to validate fluence correction algorithms and explore additional blood parameters such as pH, and also absorptive and other properties of different fluids.
\end{abstract}


{\noindent \footnotesize\textbf{*}Joanna Brunker,  \linkable{jb2014@cam.ac.uk} }

\begin{spacing}{2}   

\section{Introduction}
\label{sect:intro}  

Photoacoustic tomography (PAT) exploits optically generated ultrasound to provide images that combine the high contrast and spectral specificity of optical imaging with the high spatial resolution of ultrasound. In particular, PAT has been widely used to image blood hemoglobin concentration and oxygenation, which have the potential to inform on a range of pathophysiologies, from tumour aggressiveness \cite{MartinhoCosta2019} and treatment response \cite{Quiros-Gonzalez2018,Yang2019} to intestinal inflammation associated with Crohn's disease \cite{Waldner2016} and colitis \cite{Bhutiani2017a}. The derivation of such images is usually based on spectral unmixing to resolve the differential absorption contributions of oxy- (HbO$_2$) and deoxy-hemoglobin (Hb). Total hemoglobin concentration (THb) is typically taken as the sum of the contributions to the photoacoustic signal $\Prl$ from HbO$_2$ and Hb, while hemoglobin oxygenation (sO$_2$) is taken as the ratio of HbO$_2$ to THb. 

Unfortunately, the assessment of HbO$_2$ and Hb content from photoacoustic data is not trivial. Estimation of these chromophore concentrations from images taken at multiple wavelengths is commonly achieved using least-squares fitting of reference HbO$_2$ and Hb spectra \cite{Zhang2007b}, which are tabulated in the literature \cite{Prahl1999a,Prahl2017,Landsman1976b}; however, these spectra are recorded under \textit{in vitro} conditions thus can vary substantially from those experienced within an \textit{in vivo} study. Moreover, the measured photoacoustic signal $\Prl$ is not directly proportional to the absorbed energy density $\Hrl$. $\Hrl$ is the product of the light fluence $\Phi$, which itself varies as a function of $\mathbf{r}$ and $\lambda$, as well as $\muarl$ and the reduced scattering coefficient $\musrl$:
\begin{equation}
\Hrl = \muarl\Phirl.                   
\label{eq:absorbedEnergy}
\end{equation}
This co-dependence of $\Hrl$ on both absorption and light fluence leads to an effect known as ``spectral coloring'', where variations in local fluence bias the measured optical absorption distribution \cite{Maslov2007a,Cox2012a}. Despite this challenge, the amplitude and temporal profiles of photoacoustic waves have been shown to be linearly dependent on the blood sO$_2$ \textit{in vitro} \cite{Esenaliev2002d}. In theory, only two wavelengths are necessary to calculate two concentrations, but practical limitations mean the choice and number of wavelengths becomes critical for accurate sO$_2$ measurements \cite{Hochuli2015}. Alternative approaches beyond the standard linear unmixing model have been shown to improve sO$_2$ measurement accuracy, in particular by accounting for the absorbed energy distribution through: internal irradiation \cite{Mitcham2017}; diffusion theory modelling \cite{Laufer2005a,Cox2006c,Yuan2006c,Laufer2007b,Laufer2010a}; Monte Carlo simulations \cite{Liu2016i}; model-based iterative minimisation \cite{Yao2009a,Brochu2017,Cox2009a}; and linear superposition of reference fluence base spectra \cite{Tzoumas2016}. 

Several studies have explored ways to validate photoacoustic images using phantoms with well-characterised optical and acoustic properties \cite{Vogt2016a,Maneas2018d}. Validation of photoacoustically measured sO$_2$ is possible using CO-oximetry \cite{Esenaliev2002d,Laufer2005a,Vogt2019}, pulse oximetry \cite{Hennen2015b} and blood-gas analysis \cite{Chen2012k}; correlation between pO$_2$ and sO$_2$ has also been used to study oxygen-hemoglobin binding \cite{Wang2011s}. However, most studies have worked with static blood samples with limited control on the sO$_2$. Allowing blood to flow in a circuit \cite{Vogt2019} and with the ability to vary the sO$_2$ provides a more versatile platform for investigating PAT oxygenation measurements. 

In this work, we created a low-cost flow phantom system with online monitoring to facilitate validation of PAT systems. The phantom was tested by circulating different concentrations of methylene blue (MB) and indocyanine green (ICG). The phantom was then applied to explore the accuracy of sO$_2$ assessment using PAT, with sO$_2$ values ranging from \SI{0}{\percent} to \SI{100}{\percent}. We studied the impact of evaluating sO$_2$ using different reference spectra for HbO$_2$ and Hb, and also the effect of spectral coloring, showing discrepancies of up to \SI{60}{\percent} between the actual and measured sO$_2$. Our findings highlight the importance of careful choice of spectra for unmixing and the development of fluence correction models to improve the biological relevance of sO$_2$ measurements derived from PAT images.


\section{Methods}

\subsection{Flow system}
\label{sec:flowsytem}
\begin{figure}[h]
\centering\includegraphics[width=\textwidth]{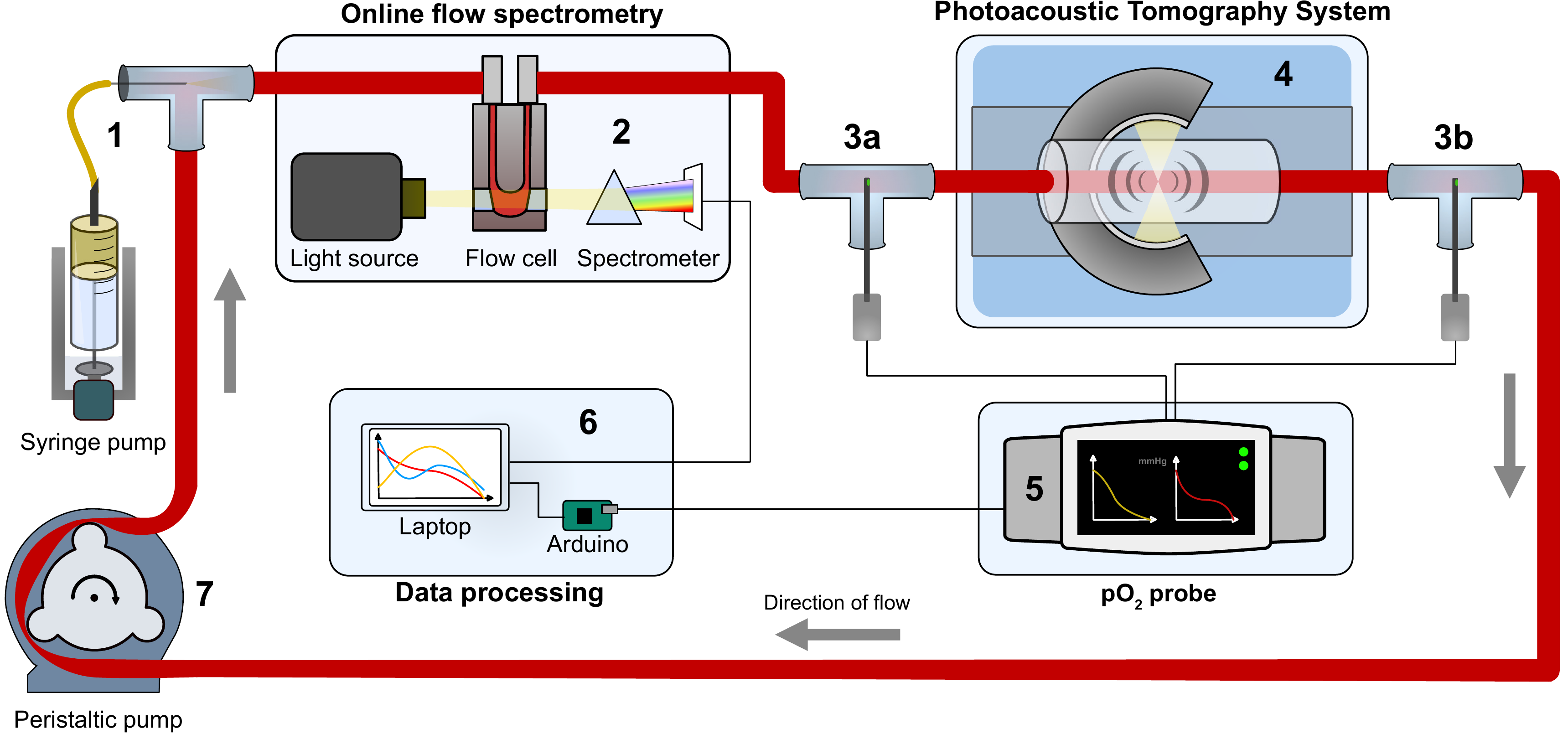}
\caption{\textbf{Overview of the flow system. (1)} Injection site for introducing oxygenated blood (or other fluids) into the flow system, and for subsequently deoxygenating the blood using sodium hydrosulfite delivered via the syringe driver (MKCB2159V, Harvard); \textbf{(2)} spectra are recorded using a light source (Avalight-HAL-S-Mini, Avantes) and spectrometer (AvaSpec-ULS2048-USB2-VA-50, Avantes) as the blood passes through a flow cell (170700-0.5-40, Hellma Analytics); \textbf{(3)} needle probes (NX-BF/O/E, Oxford Optronix) measure the temperature and partial pressure of oxygen (pO$_2$) before \textbf{(3a)} and after \textbf{(3b)} the blood passes through an agar phantom immersed in the photoacoustic imaging system (MSOT inVision 256-TF, iThera Medical) \textbf{(4)}; \textbf{(5)} a touch-screen monitor (OxyLite Pro, Oxford Optronix) displays temperature and oxygen data; \textbf{(6)} these data are downloaded via an Arduino UNO and read in MATLAB on a laptop, which also records the spectrometer readings via AvaSoft software; \textbf{(7)} a peristaltic pump (CTP100, Fisher Scientific) provides blood circulation.}
\label{fig:setup}
\end{figure}

The flow system (Fig. \ref{fig:setup}) enables fluids to be circulated within a vessel-mimicking tube embedded in a tissue-mimicking agar phantom placed in the chamber of a photoacoustic imaging system. Online (and offline) flow spectrometry and pO$_2$ detection provide independent validation of the spectral measurements made by the imaging system. Details of the tissue-mimicking phantom, the photoacoustic imaging system, the spectrometric validation, the pO$_2$ detection, and the spectral analysis are described in the following subsections.

\subsubsection{Tissue-mimicking phantom}
\label{ssc:tm-phantom}
The tissue-mimicking phantom used for photoacoustic imaging comprised either static or flowing fluids within a tube (``vessel") embedded within an agar (``tissue") cylinder, \SI{19}{\milli\meter} in diameter. Agar phantoms were prepared by heating a solution of 1.5\% w/v agar (05039, Fluka) in water, and then adding 2.1\% v/v Intralipid (I141, Sigma-Aldrich), before pouring into an open \SI{20}{\milli\liter} syringe (with the injection end removed), in the centre of which was positioned a needle supporting the tubing vertically. The solidified agar phantom had a scattering coefficient of \SI{5}{\per\centi\meter} due to the Intralipid, but the absorption coefficient was assumed to be negligible.

To provide optical absorption for assessment of spectral coloring, nigrosin (198285, Sigma-Aldrich) was also added to the tissue-mimicking phantom formulation to produce phantoms with an absorption coefficient of \SI{0.05}{\per\centi\meter} or \SI{0.1}{\per\centi\meter} at \SI{564}{\nano\meter} (the peak of the nigrosin spectrum). 

\subsubsection{Photoacoustic tomography (PAT)}
For photoacoustic imaging, a small animal imaging system (MultiSpectral Optoacoustic Tomography (MSOT) inVision 256-TF, iThera Medical) was used. Briefly, a tunable optical parametric oscillator (OPO) pumped by an Nd:YAG laser provides excitation pulses with a duration of \SI{9}{\nano\second} at wavelengths from \SI{660}{\nano\meter} to \SI{1200}{\nano\meter} at a repetition rate of \SI{10}{\hertz} with a wavelength tuning speed of \SI{10}{\milli\second} and a peak pulse energy of \SI{90}{\milli\joule} at \SI{720}{\nano\meter}. Ten arms of a fibre bundle provide near-uniform illumination over a disk extending approximately \SI{8}{\milli\meter} along the imaging chamber. Photoacoustic signals are detected using 256 toroidally focused ultrasound transducers with a centre frequency of \SI{5}{\mega\hertz} (60\% bandwidth), organized in a concave array of \SI{270}{\degree} angular coverage and a radius of curvature of \SI{4}{\centi\meter}.

\subsubsection{Detection of pO$_2$}
Two oxygen fluorescence quenching needle probes (NX-BF/O/E, Oxford Optronix) were inserted into the flow circuit before and after the tissue-mimicking phantom. A touch-screen monitor (OxyLite Pro, Oxford Optronix) displayed the temperature and partial pressure of oxygen (pO$_2$) real-time, and these data were downloaded via an Arduino UNO and read in MATLAB. The relationship between the oxygen saturation (sO$_2$) and the partial pressure (pO$_2$) in blood is described by the characteristic sigmoid-shaped oxygen-hemoglobin dissociation curve. A widely accepted fit to this curve is given by the Severinghaus equation \cite{Severinghaus1979,Collins2015}: 
\begin{equation}
\text{sO}_2 (\%) = \frac{100}{\left[23400 \times (\text{pO}_2)^3 + 150 \times (\text{pO}_2)\right]^{-1} + 1},
\label{eq:Severinghaus}
\end{equation}
which was used to convert our pO$_2$ measurements into sO$_2$.

\subsubsection{Online flow spectrometry}
Absorption spectra were recorded continuously via AvaSoft software using a light source (Avalight-HAL-S-Mini, Avantes) and spectrometer (AvaSpec-ULS2048-USB2-VA-50, Avantes) as the fluid passed through a flow cell (170700-0.5-40, Hellma Analytics). 

\subsubsection{Offline spectrometry}
To independently validate spectrophotometric measurements made in the flow circuit, optical absorbance spectra were also recorded using a microplate spectrometer (CLARIOstar, BMG LABTECH). Fluid samples were measured in a 96-well plate (Corning Costar).

\subsection{Flow system characterization}
\label{sec:flowsytemcharacterisation}

\subsubsection{Tubing assessment}
The optimum tubing was determined by comparing photoacoustic images obtained of agar phantoms containing various tube types filled with a \SI{25}{\micro\mol} ICG solution (Sigma-Aldrich I2633). The tubes were labelled according to their nominal inner and outer diameters (I.D / O.D.) in \SI{}{\micro\meter}, but were also made of different materials: polypropene 2660/2800 (Alliance Online PSTS0007); THV500 2800/3150, 500/600 (Paradigm Optics); silicone 1570/2410, 630/1190, 300/630 (VWR 228-0256, 228-0254, 228-0253); PVC 1500/2100 (VWR 228-3857); PMMA 667/1000, 432/865, 375/500 (Paradigm Optics); polythene 580/960 (Smiths Medical 800/100/200 12665497).

\subsubsection{Dye dilution series}
A dilution series was used to demonstrate the utility of the flow circuit for injecting different fluids into the closed PAT system, allowing PAT to be performed concurrently with online spectrometry. Methylene blue (MB, 50484, Fluka) and indocyanine green (ICG, I2633, Sigma-Aldrich) solutions were prepared by diluting concentrations of \SI{500}{\micro\molar} and \SI{100}{\micro\molar} (for MB and ICG respectively) in deionized water. First, four spectra were measured and averaged for samples of each concentration placed in a 96-well plate in the CLARIOstar spectrophotometer. The concentrations were then flushed individually through the flow system, starting with deionized water and then sequentially with increasing concentration. Ten online spectra were recorded over a range of \SI{333}{\nano\meter} to \SI{1100}{\nano\meter}, and ten single-slice PA images (no pulse-to-pulse averaging) were acquired for 17 wavelengths (\SI{660}{\nano\meter}, \SI{664}{\nano\meter}, \SI{680}{\nano\meter}, \SI{684}{\nano\meter}, \SI{694}{\nano\meter}, \SI{700}{\nano\meter}, \SI{708}{\nano\meter}, \SI{715}{\nano\meter}, \SI{730}{\nano\meter}, \SI{735}{\nano\meter}, \SI{760}{\nano\meter}, \SI{770}{\nano\meter}, \SI{775}{\nano\meter}, \SI{779}{\nano\meter}, \SI{800}{\nano\meter}, \SI{850}{\nano\meter}, \SI{950}{\nano\meter}), taking the mean of the ten single-slice images for each concentration. The circuit was flushed with water between each concentration.

\subsubsection{Dynamic concentration change}
A dynamic concentration change was used to illustrate the possibility for real-time spectroscopic and photoacoustic measurements. Continuous acquisition of online spectra and single-slice photoacoustic images was commenced once deionized water was circulating within the flow system. After a certain time, a high concentration dye solution (either \SI{500}{\micro\molar} MB or \SI{100}{\micro\molar} ICG) was injected using the syringe pump at (\SI{100}{\micro\liter\per\min}).

The amount (moles) of the dye $Q(t)$ in the circuit at any given time $t$ can be modelled using a first order differential equation. In Eq. (\ref{eq:DE}), the rate of change $dQ(t)/dt$ is equal to the difference between the inflow and outflow amounts, expressed in terms of the flow rate $F$ induced by the syringe pump (\SI{100}{\micro\liter\per\min}), the concentration $c$ of dye injected (\SI{500}{\micro\molar} or \SI{100}{\micro\molar}) and the volume $V$ (\SI{5}{\milli\liter}) of circulating fluid:

\begin{equation}
\frac{dQ(t)}{dt} = F\left[c - \frac{Q(t)}{V}\right].
\label{eq:DE}
\end{equation}

The initial condition is given by Eq. (\ref{eq:initial-condition}), giving rise to Eq. (\ref{eq:solution}), which is the solution for the dye concentration $Q/V$ as a function of time. In the limit $t \rightarrow \infty$, $Q(t)/V \rightarrow c$, as expected. 

\begin{equation}
Q(0) = Q_0 = 0;
\label{eq:initial-condition}
\end{equation}

\begin{equation}
\frac{Q(t)}{V} = c\left[1 - e^{-\frac{F}{V}t}\right].
\label{eq:solution}
\end{equation}

\subsection{Blood oxygenation measurements}
\label{sec:blood}

Mouse blood, predominantly made up of strains B6 and 129SvEv, was collected post mortem from the animal facility at the Cancer Research UK Cambridge Institute, complying with the UK Animals (Scientific Procedures) Act 1986. Each mouse contributed about \SI{1}{\milli\liter} to a pool of about \SI{5}{\milli\liter} blood required for the flow circuit. Chemicals were added to preserve, oxygenate and deoxygenate the blood. Prior to the experiment, ethylenediaminetetraacetic acid (EDTA) anti-coagulant (9002-07-7, Sigma-Aldrich) was added to the fresh blood, which was kept in a refrigerator under \SI{4}{\celsius} for no more than 72 hours. During the experiment, the blood was returned to room temperature, and the oxygenation was controlled chemically by adding \SI{0.2}{\percent} v/v hydrogen peroxide, H$_2$O$_2$ (7722-84-1, Sigma-Aldrich), for oxygenation \cite{WHITE1964} and sodium hydrosulfite (7775-14-6, ACROS Organics) for deoxygenation \cite{Briely-Sabo2000}. The sodium hydrosulfite was dissolved in phosphate-buffered saline (PBS) \SI{\sim0.03}{\percent} w/v for injection into the flow circuit. 

\subsection{Spectral and statistical analysis}
\label{sec:spectral_unmixing}
PAT contrast is provided by optical absorbers within the field of illumination and detection. In general, PAT does not have sufficient resolution to visualize individual molecules and therefore each image pixel (or voxel) corresponds to more than one optical absorber, and a ``spectral unmixing" approach is required in order to extract the individual spectral components. The linear mixture model \cite{Tzoumas2014a} assumes that the measured spectrum is a linear combination of distinct spectra:
\begin{equation}
    \textbf{x} = \textbf{SC} + n,
\label{eq:linear-combination}
\end{equation}
where \textbf{x} is the $M$-wavelength $\times$ $N$-pixels measurement matrix, \textbf{S} is the $M \times K$ matrix of $K$ spectra (``end-members"), \textbf{C} is the unknown $K \times N$ matrix of end-member abundances (concentrations), and $n$ is measurement noise. Neglecting $n$, an estimation $\textbf{\^{C}}$ of the absorber concentrations can be calculated from Eq. (\ref{eq:linear-combination}) by solving the following least-squares problem \cite{Cox2012a,Ding2017a}:
\begin{equation}
    \textbf{\^{C}} = \underset{C}{\arg\min}\left\lVert \textbf{SC}-\textbf{x}\right\rVert_2^2,
\label{eq:least-squares}
\end{equation}
with the solution
\begin{equation}
    \textbf{\^{C}} = \textbf{xS}^{+},
\label{eq:unmixing}
\end{equation}
where $\textbf{S}^{+}$ is the pseudoinverse of $\textbf{S}$.
The performance of spectral unmixing carried out on acquired PA images was compared using three different spectra: those from the literature \cite{Prahl1999a,Prahl2017,Landsman1976b}, and those experimentally measured in this study from either the online spectrometer cell or the offline plate reader. In addition, the online flow spectra were used to provide an independent ground truth for sO$_2$; this entailed unmixing of the online flow spectra with the spectra for oxy- and deoxy- hemoglobin also measured live during the experiments.

To correct for the spectrally varying fluence (``spectral coloring'') when using optically absorbing phantoms, the image intensities in the tube region were divided by the nigrosin absorption spectrum. Specifically, the nigrosin absorption spectrum was normalised to the known absorption coefficient (at the \SI{564}{\nano\meter} peak) and re-sampled at the 17 experimental wavelengths; the multispectral PA data were then divided by a wavelength-specific factor calculated from exponential decay over the \SI{8.5}{\milli\meter} distance (the background agar material impregnated with absorbing nigrosin) from the outer edge of the phantom to the tube wall. All data and source code used in this publication is available on \url{https://doi.org/10.17863/CAM.40365} and GitHub (\url{https://github.com/9xg/flow-phantom}).

\section{Results}

\subsection{Flow system characterization}
\label{sec:resultsCharacterisation}

\begin{figure}[h]
\centering\includegraphics[width=\textwidth]{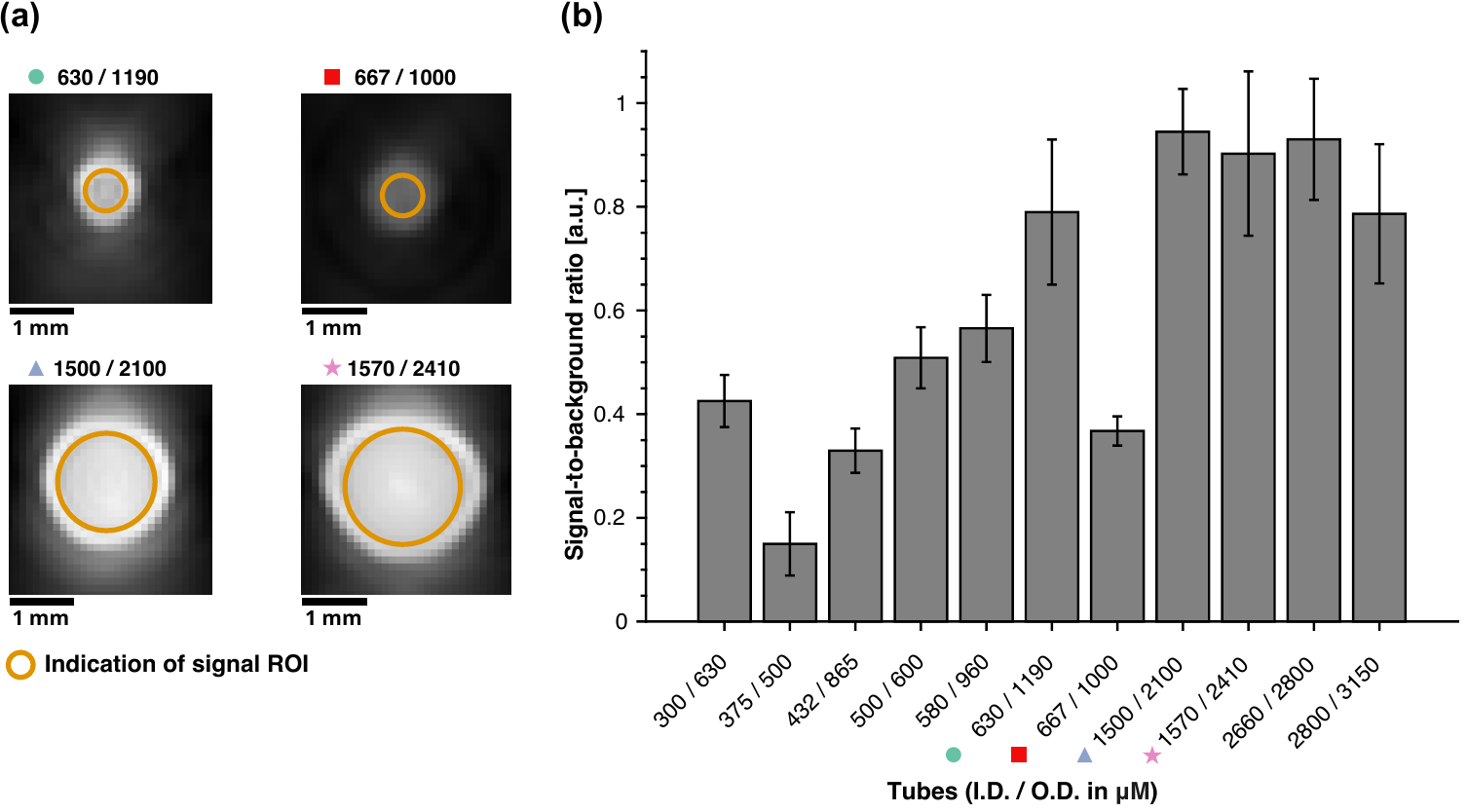}
\caption{\textbf{Assessment of the effect of tube type on the quality of photoacoustic images. (a)} PAT images of four tubes with inner and outer diameters (I.D.-O.D.) in \SI{}{\micro\meter} of 630-1190 (silicone), 667-1000 (PMMA), 1500-2100 (PVC) and 1570-2410 (silicone), filled with a solution of ICG and embedded within a scattering agar cylinder (not shown). \textbf{(b)} Mean photoacoustic signal intensity inside the tube relative to the signal outside (signal-to-background ratio, SBR) for eleven different tubes. The tube with I.D.-O.D. \SI{1500}{}-\SI{2100}{\micro\meter} was selected due to its high SBR, flexibility and low cost. The four tubes illustrated in (a) are marked with symbols.}
\label{fig:tubes}
\end{figure}

Eleven different tubes were assessed for their photoacoustic imaging suitability (Fig. \ref{fig:tubes}). 
Tubes made from PMMA (I.D./O.D. 375/500, 432/865, 667/1000) showed the lowest signal-to-background (SBR) ratios ranging from 0.15 to 0.37. The other tubes materials show a clear trend of increasing SBR with larger inner diameters. Silicone tubes present high SBRs, even for low I.D./O.D. ratios (300/600, 630/1190, 1570/2410). Three PVC tubes showed very similar performance (I.D./O.D. 1500/2100, 1570/2410, and 2660/2800). The PVC tube with 1500/2100 presented a uniform, circular appearance in images with high SBR. It was therefore selected for the remaining experiments.

\begin{figure}[h]
\centering\includegraphics[width=0.8\textwidth]{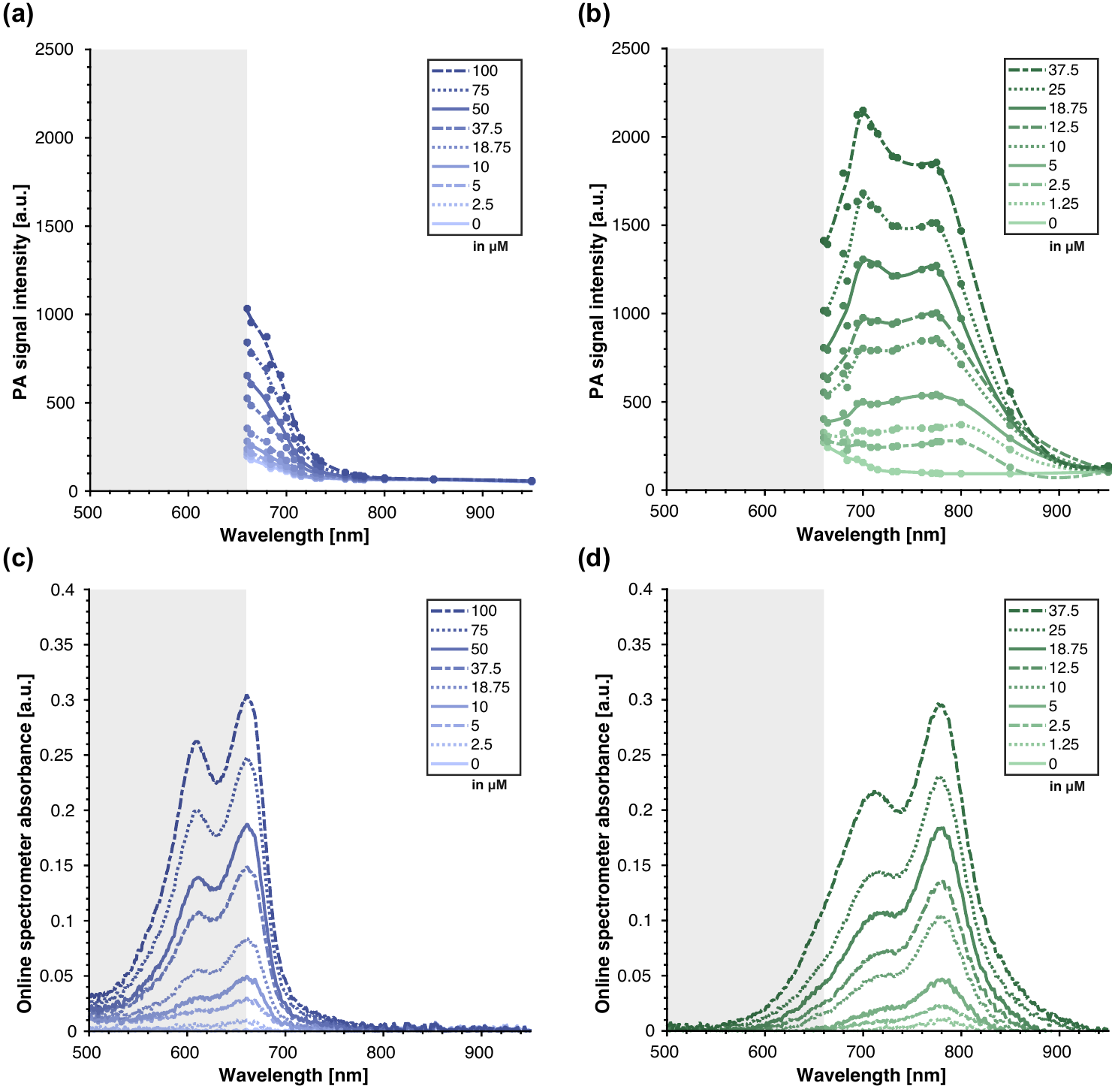}
\caption{\textbf{Comparison of dye spectra measured using PAT and online flow spectrometry under closed conditions. (a,b)} PA signal intensities (with smoothed spline) and \textbf{(c,d)} online flow spectrometer absorbance values measured for different concentrations of MB  (\textbf{a,c}), and ICG (\textbf{b,d}). Grey box indicates wavelengths outside the PAT spectral range.}
\label{fig:dye_spectra}
\end{figure}

Dilutions of MB and ICG were tested inside the flow phantom under closed conditions i.e. with a steady concentration flowing in the circuit. By calculating the mean PA signal intensities over a manually segmented tube cross-sectional region of interest (ROI) for 17 different wavelengths it was possible to compare the PA spectra with those acquired during online flow spectrometry (Fig. \ref{fig:dye_spectra} (a-d)). Unexpectedly, the two independent spectral measurements show poor agreement in terms of relative intensities and spectral shape for the different dye concentrations. In particular, it is notable that the PA signal intensities (Fig. \ref{fig:dye_spectra}b) show a more pronounced shift in spectral peak with increasing ICG concentration compared with the online spectrometer data (Fig. \ref{fig:dye_spectra}d); this may be a result of spectral coloring where light attenuation at the \SI{800}{\nano\meter} peak suppresses the PA signal intensity deeper within the tube, leading to an overall reduction in mean intensity. For MB the discrepancies are difficult to discern seeing as the prominent spectral features occur below the PA imaging range, but it is interesting that the relative PA intensities are about half those for ICG suggesting that MB has poorer PA signal generation efficiency. Overall, these spectral inconsistencies raise a question about which are the most suitable endmember spectra for unmixing PAT images of MB and ICG.

\begin{figure}[h]
\centering\includegraphics[width=0.7\textwidth]{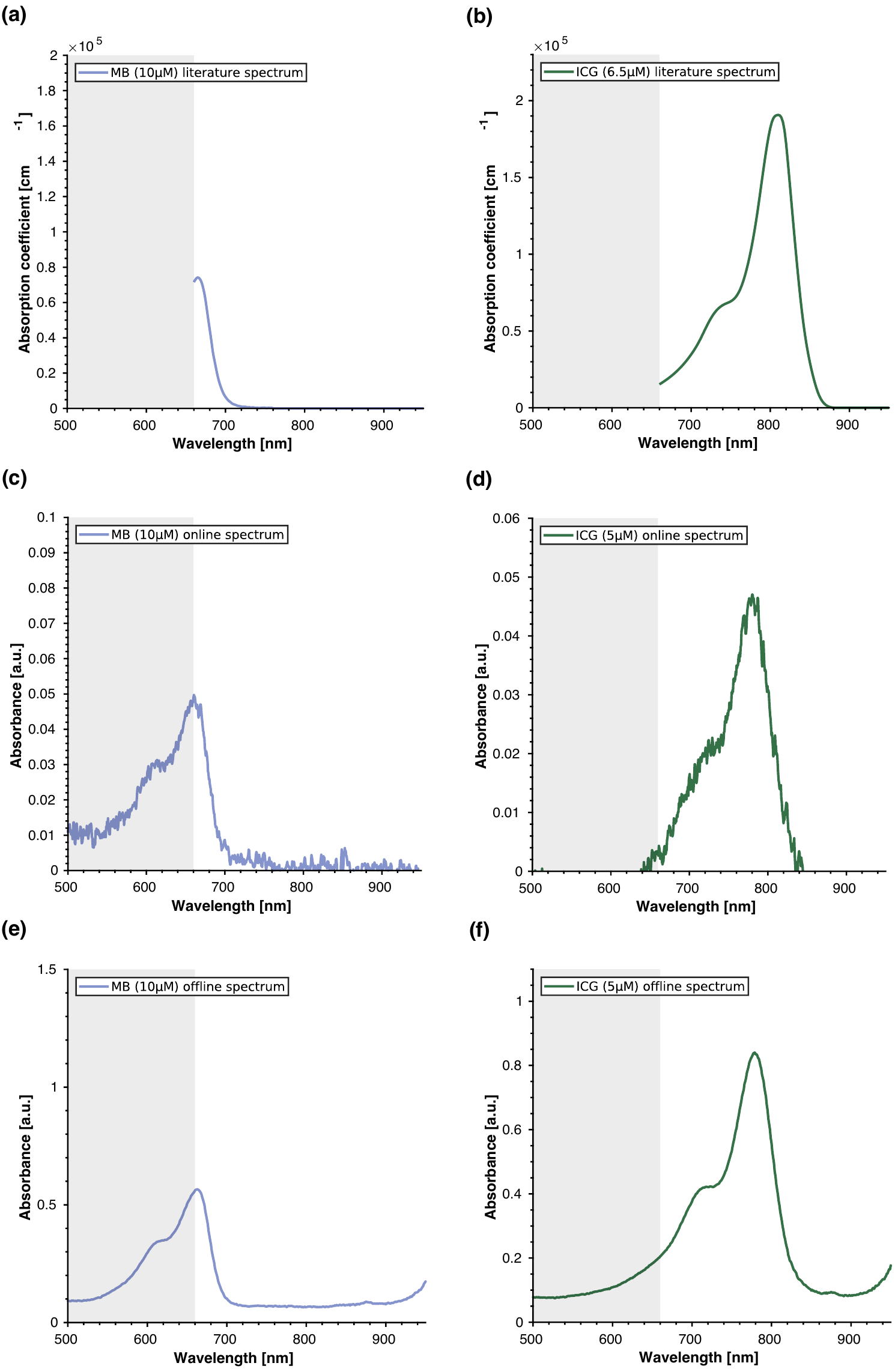}
\caption{\textbf{Comparison of endmember spectra used for spectral unmixing.} Literature spectra \textbf{(a,b)}, spectra acquired using the online flow spectrometer \textbf{(c,d)} and the offline flow spectrometer \textbf{(e,f)} for MB \textbf{(a,c,e)} and ICG \textbf{(b,d,f)}.}
\label{fig:spectra-comparison}
\end{figure}

The spectra measured with the online spectrometer and offline in the plate reader were used to perform concentration-specific unmixing of the mean ROI intensities, and these unmixed intensities were compared with those calculated using the literature spectra incorporated within the PA analysis software. These literature spectra are shown in Fig. \ref{fig:spectra-comparison} (a-b) along with online and offline spectra for example concentrations (Fig. \ref{fig:spectra-comparison} (c-f)).

\begin{figure}[h]
\centering\includegraphics[width=\textwidth]{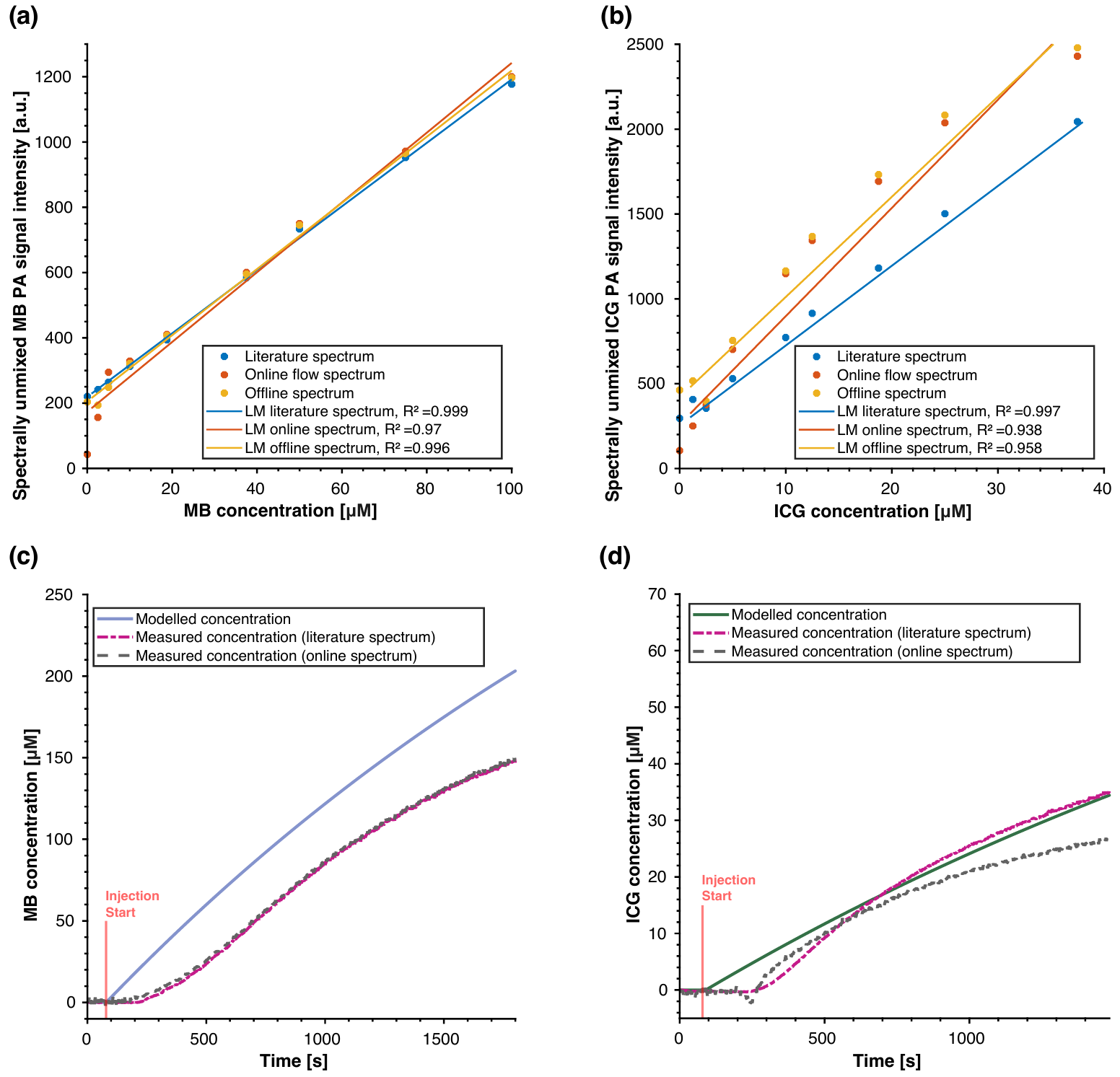}
\caption{\textbf{Spectrally unmixed PA signal intensities for a range of dye concentrations} \textbf{(a)} MB and \textbf{(b)} ICG concentrations obtained by unmixing with literature spectra and those measured using the online and offline spectrometers. The linear model (LM) was used to calculate concentrations of MB \textbf{(c)} and ICG \textbf{(d)} during evolution of dye concentrations within the flow circuit due to continual dye injection. The starting points were forced to zero for ease of comparison.} 
\label{fig:dye_linearity}
\end{figure}

Figure \ref{fig:dye_linearity} (a-b) shows a linear relationship between dye concentration and spectrally unmixed PA
intensities up to \SI{100}{\micro\mol} for MB, and \SI{40}{\micro\mol} for ICG, respectively. Divergence between the two types of unmixing is likely due to concentration dependent changes in the spectra of these dyes, which are particularly prominent for ICG.

Having explored the application of the flow circuit under static concentration values, we then examined the response of the PAT instrument to dynamic changes in dye concentrations for MB and ICG (Fig. \ref{fig:dye_linearity} (c-d)). PA images were unmixed using the literature spectra and those recorded live using the online spectrometer, and then converted to absolute concentration values using the linear models calculated and plotted in (a-b). Neglecting the initial lag phase (which could not be experimentally determined and therefore was not incorporated into the model), the rates of concentration change match reasonably well with those predicted by the model (Eq. \ref{eq:solution}). However, there is a notable discrepancy between the dynamics calculated using the literature and online spectra for ICG unmixing; if the lag phase were corrected for, the online unmixing would be the closer match to the model and this is again likely to be a consequence of the concentration-dependent change in spectral shape, which is incorporated in the online spectra but not the single literature spectrum.

\newpage
\subsection{Blood oxygenation measurements}
\label{sec:resultsBlood}

\begin{figure}[h]
\centering\includegraphics[width=0.5\textwidth]{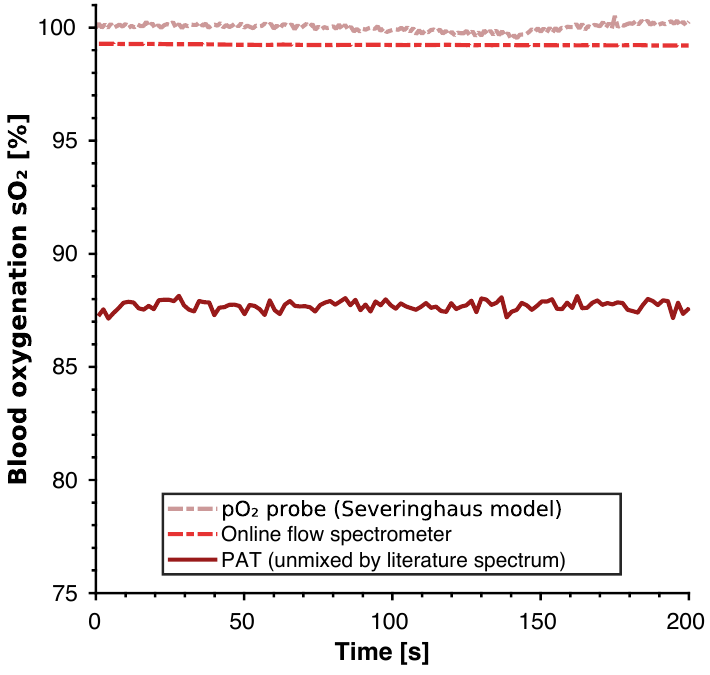}
\caption{\textbf{Assessment of blood oxygenation within the flow circuit under closed conditions.} Blood oxygen saturation (sO$_2$) was calculated using the three independent methods: the pO$_2$ probe measurements (using Equation \ref{eq:Severinghaus}); unmixing of the spectra measured using the online flow spectrometer (with experimentally measured spectra as end-members); unmixing of the mean pixel intensities in the PA images (with literature spectra as end-members) acquired while blood circulated in the flow system for 200 seconds.}
\label{fig:blood_constant}
\end{figure}

To demonstrate the closed nature of the blood flow circuit, blood was fully oxygenated by adding approximately \SI{15}{\micro\liter} \SI{0.2}{\percent} v/v H$_2$O$_2$ to \SI{8}{\milli\liter} mouse blood and then injecting this into the flow system and circulating it for several minutes. Figure \ref{fig:blood_constant} shows the blood oxygen saturation (sO$_2$) measured within the flow circuit by: the pO$_2$ probe; the online spectrometer and PAT. The sO$_2$ was calculated from the pO$_2$ data using Equation \ref{eq:Severinghaus} and from the online spectrometer and PAT data through spectral unmixing. To enable comparison between the methods independently, unmixing was performed using the spectra inherent to each system: for the online spectrometer, experimentally measured spectra were used as endmembers, while for PAT data, the literature spectra were used.  As hoped, all three sO$_2$ measurements show consistent values over time. However, the sO$_2$ calculated from the unmixed PA images under-reads by about \SI{13}{\percent} compared to the ground truth sO$_2$ calculated from the pO$_2$ probe and the online flow spectrometer.

\begin{figure}[h]
\centering\includegraphics[width=\textwidth]{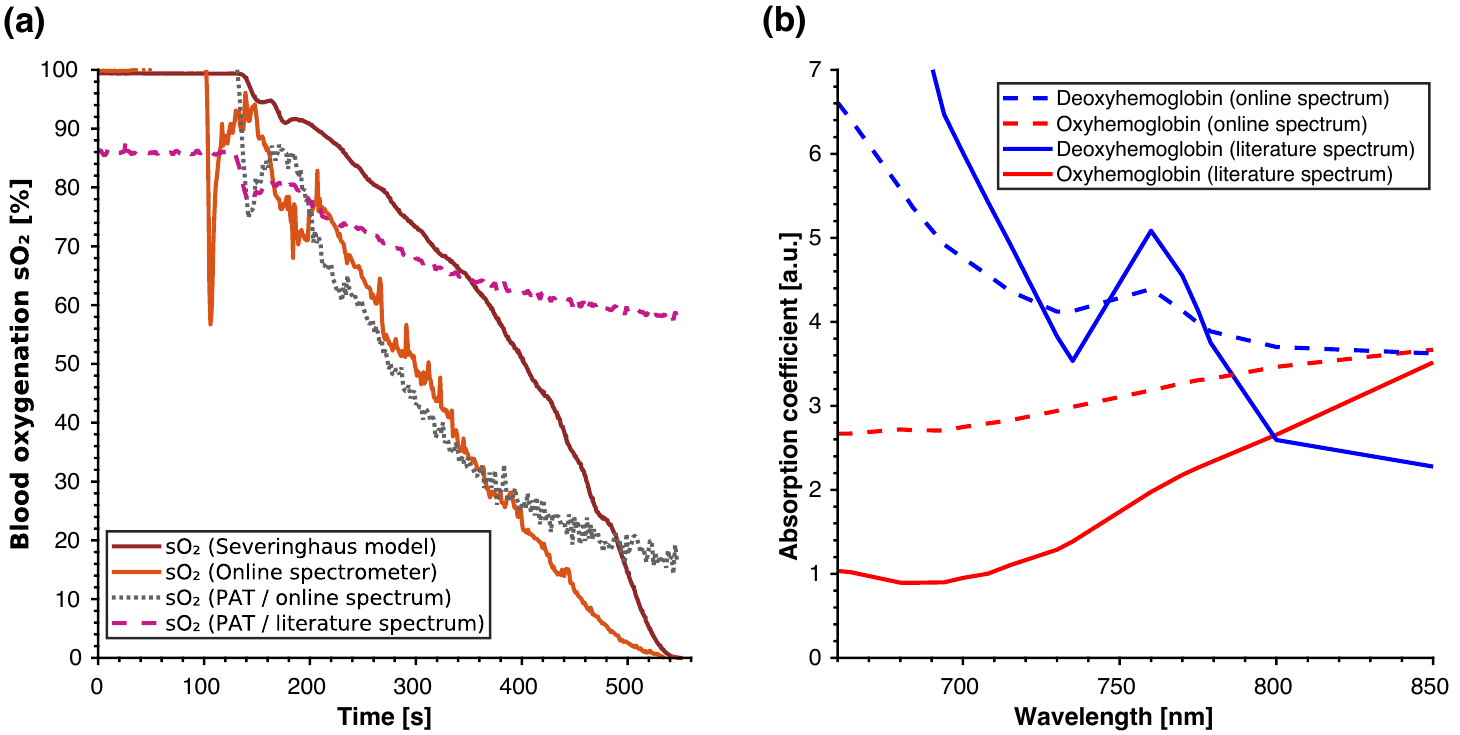}
\caption{\textbf{Dynamic deoxygenation of the circulating blood. (a)} Change in blood oxygen saturation (sO$_2$) measured using the pO$_2$ probe, online spectrometer and PAT system while injecting 3\% w/v sodium hydrosulfite in PBS over a period of nine minutes. \textbf{(b)} Literature spectra for oxygenated (HbO$_2$) and deoxygenated (Hb) hemoglobin used for spectral unmixing \cite{Prahl1999a} compared to spectra obtained from the online spectrometer at the start and end of the dynamic deoxygenation. }
\label{fig:blood_dynamic}
\end{figure}

In addition to investigating static blood oxygenation within the circuit, the full range of blood sO$_2$ values (\SI{100}{\percent} to \SI{0}{\percent}) were explored by injecting sodium hydrosulfite into the flow circuit in order to gradually deoxygenate the blood. The blood sO$_2$ values cover the expected dynamic range when calculated from both the pO$_2$ probe and the online spectrometer (Fig. \ref{fig:blood_dynamic}a); however, PA signal intensities show markedly different behaviors when unmixed with spectra (Fig. \ref{fig:blood_dynamic}b) from the literature or from the online flow spectrometer. The sO$_2$ values calculated by unmixing with the experimentally measured spectra (red line, Fig. \ref{fig:blood_dynamic}a) show good agreement with the spectrometer-derived ground truth values (yellow dotted line, Fig. \ref{fig:blood_dynamic}a), except for somewhat over-reading the oxygenation for sO$_2$ values below about \SI{20}{\percent}. They also show a dynamic range comparable to that obtained from the pO$_2$ probe. However, the sO$_2$ values calculated by unmixing with the literature spectra surprisingly exhibit a dramatically reduced dynamic range (\SI{86}{\percent} to \SI{58}{\percent}); this may be due to the unmixing spectra and/or spectral coloring. 

\begin{figure}[h]
\centering\includegraphics[width=\textwidth]{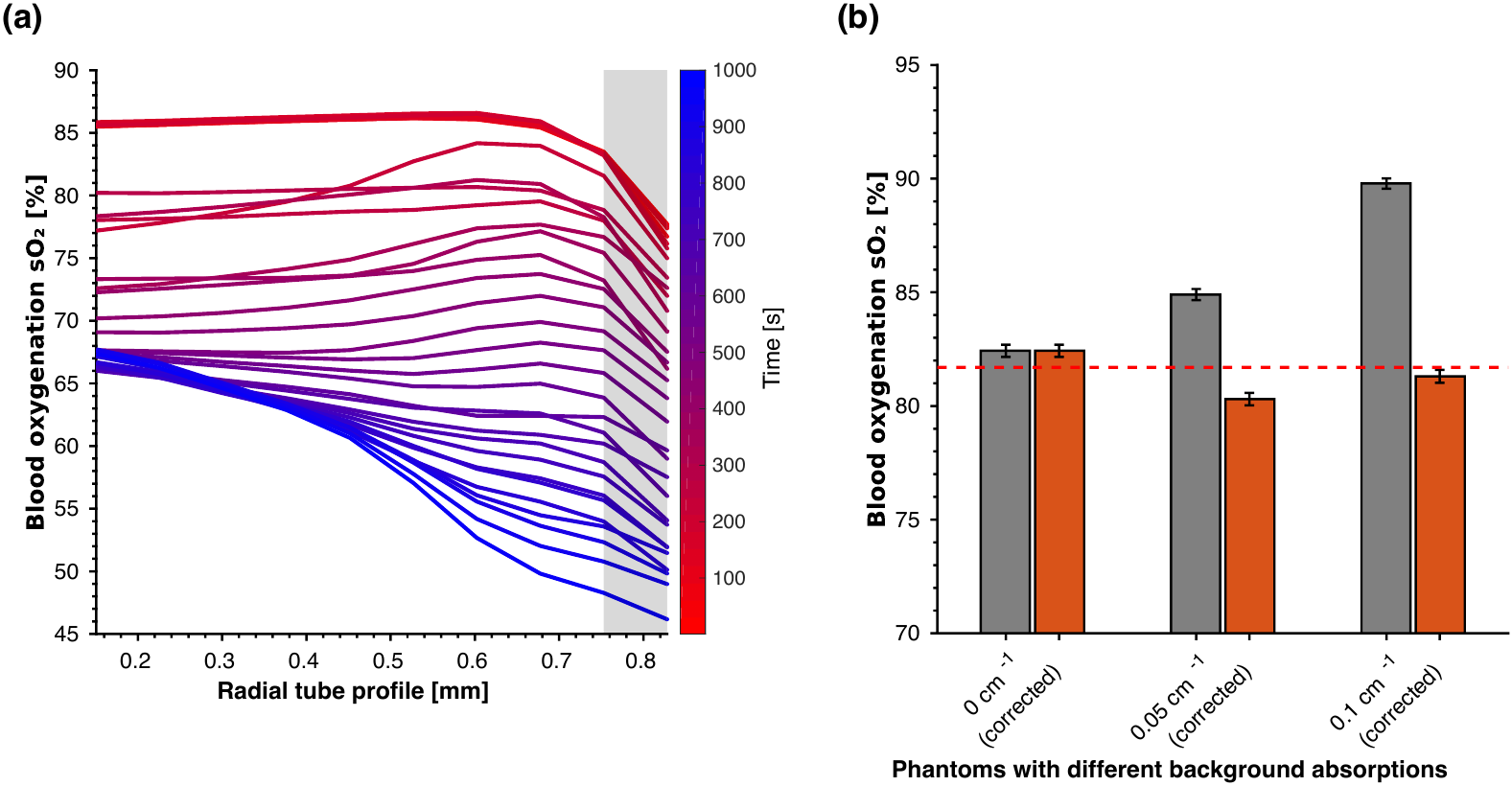}
\caption{\textbf{Investigation of spectral coloring. (a)} Spatial profiles of blood oxygen saturation (sO$_2$) as a function of radial position within the flow circuit tube by PA unmixing with spectra measured using the online spectrometer during dynamic blood oxygenation (Fig. \ref{fig:blood_dynamic}). \textbf{(b)} Impact of correction for the absorption spectrum of the nigrosin dye included in the background of the tissue mimicking phantom as a function of dye concentration. For absorption coefficients of $>\SI{0}{\per\centi\meter}$ the sO$_2$ (grey bars) deviated from the ground truth value (dotted line); accurate values were restored by recalculating sO$_2$ (orange bars) after dividing the photoacoustic images by the relevant nigrosin spectrum as described in the text.}
\label{fig:blood_profiles}
\end{figure}

Two further investigations explored the potential impact of spectral colouring. First, sO$_2$ measurements across the tube diameter were calculated (Fig. \ref{fig:blood_profiles}a). Unlike the results in Fig. \ref{fig:blood_constant} and Fig. \ref{fig:blood_dynamic} which were calculated from the mean of the ROI defined on the tube cross-sectional area, the results in Fig. \ref{fig:blood_profiles}a entailed radial unmixing using the mean of pixels found in a circle around the perimeter of the ROI, and repeating this for circle diameters reducing in one pixel (\SI{75}{\micro\meter}) increments. For the fully oxygenated blood the sO$_2$ values are consistent across the tube diameter, but as the blood becomes deoxygenated there is a clear trend towards sO$_2$ over-reading at the centre of the tube. This is a clear demonstration of spectral coloring. A second illustration of spectral coloring is shown in Fig. \ref{fig:blood_profiles}b where experiments were repeated with absorbing nigrosin dye incoporated into the background of the tissue mimicking phantom surrounding the flow circuit tube. Increasing the background absorption causes increasing inaccuracy of the sO$_2$ relative to the ground truth since wavelengths of light are preferentially absorbed by the nigrosin dye, leading to over-estimation of the contribution of oxyhemoglobin to the PA signal. A simple light fluence correction, implemented by dividing the images by the known nigrosin spectrum (section \ref{sec:spectral_unmixing}), restored consistent sO$_2$ measurements  irrespective of the background absorption, providing evidence that this spatial phenomenon is indeed related to spectral coloring.

\section{Discussion}
\label{sec:discussion}

The application of PAT in assessment of blood hemoglobin concentration and oxygen saturation is now widespread, but estimation of the oxy- and deoxy-hemoglobin chromophore concentrations is affected by a variety of tissue- and instrumentation-dependent factors. We have developed and applied a low-cost flow phantom system with online monitoring of optical absorption spectra and partial pressure of oxygen in order to facilitate testing of PAT systems.

We first selected an appropriate tubing to maximise PAT image quality and tested the circuit using two optically absorbing dyes. For both dyes, MB and ICG, the absorption spectrum exhibits a double peak where the relative peak intensities shift with increasing concentration. Unmixing with a single ``average'' spectrum is therefore unlikely to yield the same result as unmixing with a concentration-specific (online/offline) spectrum; indeed, we observed different and slightly non-linear relationships between dye concentration and spectrally unmixed intensities (which are assumed to be proportional to concentration) for the two types of unmixing. For ICG, the values from unmixing with the literature spectrum under-read those from unmixing with the online/offline spectra (Fig. \ref{fig:dye_linearity}b,d); for MB the unmixing results were indistinguishable (Fig. \ref{fig:dye_linearity}a,c) since the concentration dependent effects occur outside of the wavelength range of the PA system under test. It is possible that erroneous concentration measurements may also be caused by spectral coloring, even though only a single component is used in the spectral unmixing; two component unmixing (Hb and HbO$_2$) is more relevant for calculation of sO$_2$.

We proceeded to evaluate the estimation of sO$_2$ under both closed circuit and dynamic conditions. The results highlight two major factors that contribute to inaccuracies in spectral unmixing, and therefore in sO$_2$ measurements. The first is the choice of spectra for unmixing the photoacoustic images; the second is the effect of spectral coloring. The online spectrometer allowed spectra to be recorded during PAT image acquisition and used as alternatives to published spectra for unmixing the images. Moreover, the ground truth sO$_2$ values provided by the spectrometer and the pO$_2$ probes are designed to facilitate exploration of phenomena such as spectral coloring.

Images of blood flowing through the circuit entailed unmixing with two components: oxy- and deoxy- hemoglobin. The resulting sO$_2$ calculations (HbO$_2$/THb) were consistent over time but were dependent on the hemoglobin spectra used for unmixing, the blood oxygenation saturation, and also the position within the tube. The spectra for oxy- and deoxy- hemoglobin measured online were considerably different from those available in the literature that are widely used for spectral unmixing both \textit{in vitro} and \textit{in vivo}. Using our online flow spectrometer spectra improved the accuracy of the sO$_2$ calculations compared to those made using the default spectra, suggesting that the online spectra were more representative of the actual optical absorption presented to the PAT system.

Over-reading of the lowest sO$_2$ values was also observed and attributed to spectral coloring: the high absorption by deoxygenated blood at wavelengths below about \SI{750}{\nano\meter} results in low light intensities at these wavelengths in the centre of the tube and therefore the PA intensities (proportional to both absorption and light fluence) varied with wavelength in a way more closely matching with oxyhemoglobin, leading to overestimation of the oxyhemoglobin concentration. Admittedly, these sO$_2$ values are well below the physiological range (\SIrange{70}{100}{\percent}), nonetheless it is important to explore the full dynamic range in order to establish the accuracy limits of PA sO$_2$ measurements. A simple example of correcting the effects of spectal coloring introduced by the tissue mimicking phantom surrounding the tube was achieved through dividing by the background nigrosin spectrum, which restored accurate sO$_2$ measurements. 

One limitation of the study is the discrepancy between our measured hemoglobin spectra and those widely used in the literature. This could be attributed to optical scattering due to the blood cells within the measurement cuvette, as literature studies normally measure the haemoglobin molecule directly; another factor could be the strong attenuation of the blood, necessitating use of a neutral density filter in order to prevent signal saturation when recording the reference spectrum (taken with PBS in the tube), but then removal of the filter for the blood measurements. Although the filter was taken into account in the calculation of abosrption spectra, the effect of blood scattering is difficult to compensate for; future experiments should therefore calibrate the measured spectra by extracting blood samples from the circuit at different sO$_2$ values and validating the sO$_2$ using a blood gas analyser. It would also be interesting to investigate oxygenation states of haemoglobin extracted from blood cells thereby obviating optical scattering effects. A further limitation arises from the pO$_2$ measurements, as the dynamics of the deoxygenation study did not directly mirror the sO$_2$ values obtained in the spectrometer. This may be due to an unrepresentative conversion from pO$_2$ to sO$_2$: the Severinghaus equation is derived from human, not mouse, blood data and also assumes certain values for parameters such as pH and temperature. Alternative conversions such as the Kelman equation were also explored, and measured values for pH and temperature were incorporated, but this did not significantly alter the trend in sO$_2$ over time. It is possible that bubbles or insufficient contact with the blood in the circuit corrupted the pO$_2$ readings, and therefore future versions of the circuit will integrate truly in-flow pO$_2$ probes and a membrane oxygenator to avoid bubbles. 

The application of the presented phantom opens a range of opportunities for future studies of tissue- and instrument-dependent correction factors in PAT. In particular, future work will attempt to model and account for the spectral distortion across the tube diameter observed during blood oxygenation studies. The development of corrections incorporating multiple choromphores inside and outside the tube both \textit{ex vivo} and \textit{in vivo} is more complex and continues to be investigated, for example using Monte Carlo simulations and model-based iterative minimisation \cite{Cox2006c,Yuan2006c,Yao2009a,Brochu2017,Cox2009a,Laufer2010a}. Moreover, the number and choice of wavelengths used for spectral unmixing is an important consideration \cite{Hochuli2015} to be investigated in future work.

\section{Conclusion}
\label{sec:conclusion}

In summary, we have developed a low-cost flow phantom that includes an online spectrometer and partial pressure of oxygen probe to facilitate detailed validation of PAT measurements of blood oxygen saturation, sO$_2$. We found that it is important to correctly identify the absorption spectra to be used for unmixing photoacoustic images in order to accurately determine even relative absorber concentrations. We also found that the co-dependence of photoacoustic signal intensity on light fluence and absorption leads to a major challenge in accounting for spectral coloring, which can lead to substantial underestimation of sO$_2$.  Further calibration and automation of the circuit will enable additional \textit{ex vivo} studies requiring careful control and knowledge of blood sO$_2$, as well as opening the possibility of exploring additional blood parameters such as pH, and also absorptive and other properties of different fluids. Detailed understanding of the photoacoustic signal origins \textit{ex vivo} remains essential for proper interpretation of photoacoustic measurements made \textit{in vivo}.

\appendix

\subsection*{Disclosures}
SEB has received research support from iThera Medical GmbH and PreXion Inc., vendors of photoacoustic imaging instruments.

\acknowledgments 
The authors would like to thank Ayaka Shinozaki for her assistance with data collection, James Joseph for his contributions to the experimental design, and Michael Schneider for his helpful input regarding data interpretation. This work was supported by Cancer Research UK (C47594/A16267, C14303/A17197) and the EPSRC-CRUK Cancer Imaging Centre in Cambridge and Manchester (C197/A16465 and C8742/A18097). We would like to thank the CRUK CI Core Facilities for their support of this work, in particular the Imaging Core and the Biological Resource Unit.   


\bibliography{report}   
\bibliographystyle{spiejour}   


\vspace{2ex}\noindent\textbf{Marcel Gehrung} completed his MSc degree at Eberhard Karls University Tuebingen in 2017 and is presently completing his PhD in Medical Science at the University of Cambridge focused on the application of advanced image analysis methods for cancer diagnosis.

\vspace{2ex}\noindent\textbf{Sarah Bohndiek} completed her BA degree at the University of Cambridge in 2005 and her PhD in Radiation Physics at University College London in 2008, specialising in X-ray diffraction. After postdoctoral fellowships in both the UK and USA, she started the VISIONLab in Cambridge in 2013, which develops and applies new imaging biomarkers to shed light on the tumour microenvironment.

\vspace{2ex}\noindent\textbf{Joanna Brunker} received her MSc degree in natural sciences (2009) and her PhD (2013) in Medical Physics and Biomedical Engineering from the University College London, UK. Following her postdoctoral research fellowships at UCL and at the Cancer Research UK Cambridge Institute, she is now (July 2019) starting a research group at the UCL Wellcome / EPSRC Centre for Interventional and Surgical Sciences (WEISS) where she will focus on translation of photoacoustic and other imaging technologies into clinical practice.

\listoffigures

\end{spacing}
\end{document}